\documentstyle[12pt,cite,amssymb]{article}

\newcommand{\al}{\alpha}

\newcommand{\g}{\gamma}

\newcommand{\la}{\lambda}
\newcommand{\s}{\sigma}
\newcommand{\eps}{\epsilon}
\newcommand{\om}{\omega}
\newcommand{\ts}{\tilde{\sigma}}
\newcommand{\G}{\Gamma}
\newcommand{\D}{\Delta}
\newcommand{\La}{\Lambda}
\newcommand{\Si}{\Sigma}

\newcommand{\cL}{{\cal L}}

\newcommand{\cO}{{\cal O}}
\newcommand{\cA}{{\cal A}}
\newcommand{\cB}{{\cal B}}
\newcommand{\cD}{{\cal D}}
\newcommand{\beq}{\begin{equation}}
\newcommand{\eeq}{\end{equation}}
\newcommand{\bea}{\begin{eqnarray}}
\newcommand{\eea}{\end{eqnarray}}
\newcommand{\bean}{\begin{eqnarray*}}
\newcommand{\eean}{\end{eqnarray*}}
\newcommand{\non}{\nonumber}
\newcommand{\Bp}{\mbox{I}\!\mbox{P}}

\newcommand{\tW}{\widetilde{W}}
\newcommand{\dol}[2]{H_{\bar{\partial}}^{#1,#2}}

\begin{document}

\setcounter{page}{0}
\pagestyle{empty}
\vspace*{15mm}
\begin{flushright}
{CERN-TH/99-114
\\ hep--th/9904218
}
\end{flushright}
\vspace*{5mm}
\begin{center}
{\bf On the twisted chiral potential in 2d and the analogue of rigid
special geometry for 4-folds} \\
\vspace*{1cm} 
Peter Kaste$^{*}$ \\
\vspace{0.3cm}
Theoretical Physics Division, CERN \\
CH - 1211 Geneva 23 \\
\vspace*{2cm}  
{\bf Abstract} \\ \end{center}
\vspace*{5mm}
\noindent
We discuss how to obtain 
an $N=(2,2)$ supersymmetric $SU(3)$ gauge theory in two dimensions
via geometric engineering from a Calabi-Yau 4-fold and compute its 
non-perturbative twisted chiral potential $\tW(\Sigma)$. 
The relevant compact part of the 4-fold geometry consists 
of two intersecting ${\Bp}^1$'s fibered over ${\Bp}^2$.
The rigid limit of the local mirror of this geometry is a complex
surface that generalizes the Seiberg-Witten curve and on which there
exist two holomorphic 2-forms. These stem from 
the same meromorphic
2-form as derivatives w.r.t.\ the two moduli, respectively.
The middle periods of this meromorphic form 
give directly the twisted
chiral potential. The explicit computation of these 
and of the four-point Yukawa couplings allows for
a non-trivial test of the analogue of rigid special geometry
for a 4-fold with several moduli.

\vspace*{2cm} 
\noindent 
\rule[.1in]{6.5cm}{.002in}

\noindent
$^{*}$ e-mail: peter.kaste@cern.ch

\vspace*{0.5cm}

\begin{flushleft} CERN-TH/99-114 \\
April 1999
\end{flushleft}
\vfill\eject

\pagestyle{plain}

\section{Introduction}

The embedding of supersymmetric gauge theories into the framework of 
string theory and the subsequent application of symmetries and dualities
of the latter has shown to be an extremely fruitful approach to the study
of non-perturbative properties of gauge systems. 
One way to proceed is to exploit the fact that type IIA string theory gives
rise to non-abelian gauge symmetries when compactified on 
certain singular Calabi-Yau manifolds.
These are naturally $K3$-fibrations and a gauge theory 
of ADE-type arises
when the $K3$ fiber develops a corresponding singularity. ``Geometric
engineering'' \cite{kkv} furthermore exploits the fact 
that in the field theory limit the relevant part of the compactification 
geometry is the local singularity structure. This allows us, for the purpose of
extracting field theoretical properties, to just model manifolds that exhibit
the correct configuration of exceptional divisors (${\Bp}^1$'s that occur
in resolving the singularities of the $K3$ fiber and whose intersection 
matrix equals the negative of the
Cartan matrix of the respective gauge group) fibered
over the appropriate base. 

The literature on geometric engineering of $N=2$ supersymmetric gauge theories
in four dimensions is already vast. In this paper 
we follow \cite{fi} in
applying this method to Calabi-Yau 4-folds, which 
leads to $N=(2,2)$ supersymmetric 
gauge theories in two dimensions. 
The specific example we will investigate is an $SU(3)$ gauge theory.
This implies that the local compactification geometry on the type IIA side
contains two intersecting ${\Bp}^1$'s fibered over a common compact complex
two dimensional base, which we take to be a $\Bp^2$.
The rigid limit of the local mirror to this geometry is a complex surface,
which plays the r\^ole of the Seiberg-Witten curve.
In analogy to the 3-fold case, this surface is no longer 
a Calabi-Yau manifold
as it was in the $SU(2)$ example of \cite{fi}. Instead, it has
two holomorphic 2-forms that stem from the same meromorphic 2-form as the
derivatives w.r.t.\ the two moduli, respectively.   
A major novelty of 4-folds, as compared to 3-folds, is that 4-forms are no
longer dual to 2-forms but represent an
independent part of the cohomology.
In particular the primitive subspace of $\dol{2}{2}(X)$, i.e.\ the one 
generated by forms in $\dol{1}{1}(X)$, will play a predominant r\^ole. It
is related to the occurrence of 4-fluxes $\nu_a$, which as well as part of
the intersection form $\eta^{(2)}$ on this subspace
appear in the twisted chiral potential $\tW$.
It is given by
\beq
 \tW=\nu \cdot \eta^{(2)} \cdot \s_{D1} 
  \ , \label{intro:W}
\eeq
where $\s_{D1}$  
are the middle periods
of the meromorphic 2-form 
on the rigid surface.

We will explicitly compute these periods and the four-point Yukawa couplings 
and 
then perform a non-trivial test of the generalization of rigid special
geometry to 4-folds with several moduli. 
Recall that for 3-folds, $X_3$,
special geometry
implies that the prepotential $F$ 
of the complex structure moduli space
is given in terms of the periods 
$Z^i,F_j$ of the 
holomorphic 3-form w.r.t.\  a symplectic basis of $H_3(X_3,{\mathbb Z})$
as $F=\frac{1}{2}Z^i F_i(Z)$. 
The three-point Yukawa couplings are then given
by $C_{(3)}=\partial^3 F$, where derivatives are
w.r.t.\ the special projective coordinates $Z^i$. 
We will verify that for 4-folds the analogous
structure in the rigid limit is given by $\tW$ as 
in (\ref{intro:W}) and by
\[
 C_{(4)}\sim (\partial^2 \s_{D1})\cdot \eta^{(2)} \cdot 
 (\partial^2 \s_{D1}) \ .
\] 
We will make this more precise in the following. 

The computation of the four-point Yukawa couplings is 
done in a global model of the compact Calabi-Yau 4-fold, a realization
of which is given by
the resolution of the Fermat hypersurface of degree 36 in the 
weighted projective
space ${\Bp}^{5}(18,12,3,1,1,1)$, which according to common convention 
we will 
call $X_{36}(18,12,3,1,1,1)$. 
This manifold does not only represent a $K3$
fibration over ${\Bp}^2$, but the $K3$ is itself a fibration of an
algebraic 2-torus over ${\Bp}^1$.
Such elliptically fibered 4-folds are interesting in themselves as they give
rise to phenomenologically more interesting $N=1$ supersymmetric gauge theories
in four dimensions when used as compactification manifolds for $F$-theory
\cite{fth}.

\section{The holomorphic Fayet-Iliopoulos potentials}

It is well known that for generic points in the moduli space spanned by the
gauge multiplets the non-abelian gauge group of $N=2$ supersymmetric gauge
theories is broken down to the maximal abelian torus. Hence we will have to
consider an effective gauge theory with $N=(2,2)$ supersymmetry and abelian
gauge group $U(1)^k$ in two dimensions, where in our specific example $k=2$.
The superfields appearing in such theories comprise $k$ real 
vector superfields $V_a$,
$a=1,\ldots,k$, with component expansion
\begin{eqnarray*}
 V_a &=& -\sqrt{2} \left( \theta^- \bar{\theta}^- v_{\bar{z},a}+
         \theta^+ \bar{\theta}^+ v_{z,a}-\theta^- \bar{\theta}^+ \sigma_a
         -\theta^+ \bar{\theta}^- \bar{\sigma}_a \right) + \\
     & & +\, i \left( \theta^2 \bar{\theta}^{\dot{\al}} 
          \bar{\lambda}_{\dot{\al},a}
         -\bar{\theta}^2 \theta^{\al} \lambda_{\al,a} \right)
         +\frac{1}{2} \theta^2 \bar{\theta}^2 D_a \  ,
\end{eqnarray*}   
$r$ chiral superfields $\Phi_i$, $i=1,\ldots,r$, 
obeying $\bar{D}_{+} \Phi_i=\bar{D}_{-} \Phi_i=0$, with component expansion
\[
 \Phi_i = \phi_{i}+\sqrt{2} \left(\theta^+ \psi_{+,i} + \theta^- \psi_{-,i} 
  \right) + \theta^2 F_{i} + \ldots \  ,
\]
where $\ldots$ are total derivative terms, as well as their complex conjugates 
$\bar{\Phi}_i$. As a novelty of two dimensions there are in addition 
twisted chiral superfields $\Si_{a}$, $a=1,\ldots,k$, that satisfy 
$\bar{D}_{+} \Si_a=D_{-} \Si_a=0$. Their component expansion reads
\bea
 \Si_a &=& \frac{1}{\sqrt{2}} \bar{D}_{+} D_{-} V_a  \non \\
      &=& \s_a-i\sqrt{2} \left( \theta^+ \bar{\la}_{+,a} + \bar{\theta}^{-} 
 \la_{-,a} \right)
          + \sqrt{2} \theta^+ \bar{\theta}^- \left( D_a-i f_a \right) 
          + \ldots  \label{expand:sig}
\eea
The most general Lagrangian involving these superfields consists of a
generalized K\"{a}hler potential $K(\Si,\bar{\Si},\Phi,\bar{\Phi})$ as well as
holomorphic chiral and twisted chiral potentials, $W(\Phi)$ and $\tW(\Si)$.
 
For generic points in the moduli space the chiral matter fields will be
massive, so that in the infrared, after having them integrated out,
we are left with only the twisted chiral fields as the light degrees
of freedom and an effective action involving $K(\Si,\bar{\Si})$ and $\tW(\Si)$.
Taking the scaling dimension of $\Si$ 
to equal $1$, the K\"ahler potential has to
be multiplied by the squared inverse of a dimensionful gauge coupling and
therefore becomes irrelevant in the infrared. The twisted chiral potential
on the other hand generalizes the Fayet-Iliopoulos term
\beq 
 \frac{i}{2\sqrt{2}} \int d\theta^+ d\bar{\theta}^- 
  \left. \tW(\Si) \right|_{\theta^-=\bar{\theta}^+=0}
  + \mbox{c.c.} = 
 \sum_{a=1}^k \left(-\xi_{a}(\s) D_a+\frac{\theta_{a}(\s)}{2\pi} 
  f_a \right)
\eeq
and gives rise to effective, field dependent, complex FI couplings
\beq
 \tau_{a}(\s) \equiv i \xi_{a}(\s)+\frac{\theta_{a}(\s)}{2\pi} = 
  \left. \frac{\partial \tW(\Si)}{\partial 
  \Si_{a}} \right|_{\theta=\bar{\theta}=0} \ , \qquad a=1,\ldots,k \ . 
  \label{def:tau}
\eeq
These dimensionless couplings are known \cite{phases,sum} to receive
logarithmic perturbative corrections to one-loop order
\beq
 \tau_{a}(\s)=\tau_{a,0}-\frac{1}{2\pi i}\sum_{i=1}^{r} Q_{i}^{a}
   \log\left(\frac{\sqrt{2}}{\mu}\sum_{b=1}^{k} Q_{i}^{b} \s_{b}\right)
   +\ldots \ , \label{tau:RG}
\eeq
where $\mu$ is the RG scale and $Q_{i}^{a}$ is the charge of the 
$i^{\rm th}$ massive
chiral matter field under the 
$a^{\rm th}$ $U(1)$ factor. In addition we expect 
non-perturbative corrections. 

A further difference from four dimensional gauge theories is the
appearance of a non-trivial scalar potential
\beq
 V(\s) \sim \sum_{a=1}^{k} \left| \tau_{a}(\s) \right|^2 \ . \label{spot}
\eeq
This potential makes the vacuum energy depend on the theta-angle
\cite{col,phases} and implies that supersymmetry is broken 
unless $\tau_{a}(\s)=0$
for all $a=1,\ldots,k$.

\section{The twisted chiral potential via string theory}
 
In compactifications of type IIA string
theory on a  Calabi-Yau 4-fold $X$, 
the twisted chiral superfields $\Si_a$ of the resulting 
two dimensional gauge theory
are in one-to-one correspondence with K\"ahler classes in
$\dol{1}{1}(X)$, while the chiral matter fields $\Phi_{i}$ correspond to the
complex structure moduli belonging to $\dol{3}{1}(X)$. The twisted chiral
potential $\tW(\Si)$ is 
holomorphic in the $\Si$'s and is
thus a holomorphic section of a line bundle over the  moduli
space of K\"ahler deformations of $X$. Since the type II dilaton resides in
a different multiplet, $\tW(\Si)$ does not receive any quantum space-time
corrections and we can restrict ourselves to string tree level. 
Nevertheless $\tW(\Si)$ suffers from non-perturbative corrections due to
embeddings of the worldsheet ($\Bp^1$, since we are in string tree-level) into
$X$, i.e.\ from worldsheet instantons. These lead to quantum corrections of
the K\"ahler moduli space of $X$ that show up in $\tW(\Si)$. The way to
compute these corrections is to use mirror symmetry, which allows us to
consider instead type IIA string theory on the mirror manifold $X^*$, 
while the
K\"ahler moduli space of $X$ is mapped to the complex structure moduli space
of $X^*$ (isomorphic to $\dol{3}{1}(X^*)$) 
and vice versa. Here the twisted chiral
potential is a holomorphic section 
of a line bundle over the moduli space of complex structure
deformations of $X^*$; as such, it receives no quantum corrections at all.

As was done in \cite{fi}, 
we can identify the tree-level correlator we have to
compute in order to obtain $\tW(\Si)$ by considering the following tree-level
Chern-Simons term in ten dimensional type IIA string theory
\[ \cL_{\rm CS} = B \wedge F_4 \wedge F_4 \ , \]
where $F_4$ is the field strength of the RR 3-form field of type IIA theory.
We expand the above forms in topological bases  $\{\cO^{(i)}_{a_{i}}\}$ of
$\dol{i}{i}(X)$ as
\beq
 B=\sum_{a=1}^{h^{1,1}} \s_a \cO^{(1)}_a \ , \qquad
 F_4=\sum_{b=1}^{\tilde{h}^{2,2}} \nu_b \cO^{(2)}_b +
  \sum_{a=1}^{h^{1,1}} F_a \wedge \cO_a^{(1)} \ , 
\eeq
where $F_a$ is the field strength of the twisted chiral superfield, related
to the component $f_a$ in the expansion (\ref{expand:sig}) of $\Si_a$
by $F_a=f_a d^2x$, with $d^2x$ the volume form on the complement of $X$.
Actually the expansion in the space $\dol{2}{2}(X)$ is restricted to
its primitive subspace, whose dimension we denote by $\tilde{h}^{2,2}$.
Novel to 4-folds as compared to 3-folds is the r\^ole played by elements
$\cO^{(2)}_{b}$ and their corresponding coefficients $\nu_{b}$. 
Denoting the dual cycles in $H_{4}(X,{\mathbb Z})$
by $\{\g_{b}\}$, 
we have
\beq
 \int_{\g_{b}} F_{4} = \nu_{b} \ , \label{4fluxes}
\eeq
and it is known \cite{4flux} that 
at the quantum level these 4-fluxes have to be integers (or possibly 
half-integers if $\frac{p_1}{4}$ is not an integral class).
Using the above expansion we obtain
\beq
 \langle \cL_{\rm CS} \rangle_{\rm 10 d} \, \sim 
 \s_{a} n_{b} \nu_{c} \langle \cO^{(1)}_{a}
  \cO^{(1)}_{b} \cO^{(2)}_{c} \rangle_{X} \ , \label{Chern:Simons}
\eeq
where $n_b=1/(2 \pi) \int F_b$ is the first Chern class 
of the $b^{\rm th}$ $U(1)$ bundle,
i.e.\ the instanton number of the $b^{\rm th}$ gauge factor. 
Note that the correlator on the right hand side is a Yukawa coupling in a
topological sigma model, known as the A model \cite{top}, which is obtained by
twisting the superconformal sigma model on the worldsheet with target space
$X$. The algebra of observables of this model is identified with the quantum
deformation of the classical intersection algebra on $\cA=\oplus_{p=0}^{d}
H^p(X,\wedge^p T^*X)$, where $d$ is the complex dimension of $X$. 
Another twist leads to the B model, whose
algebra of observables is the algebra on $\cB=\oplus_{p=0}^{d} H^p(X,\wedge^p
TX)\cong \oplus_{p=0}^{d} H^p(X,\wedge^{d-p} T^*X)$. 
Mirror symmetry relates the
A and B models to each other on a pair of mirror manifolds . It was shown in
\cite{mird,klemm4,mayr4} how to compute the Frobenius
subalgebra of $\cA$ corresponding to the primitive part of the vertical
cohomology of $X$ around the large radius point, 
which is entirely determined in
terms of the two- and three-point functions. 
For the case of several moduli,
the authors of
\cite{klemm4,mayr4} found in particular that the couplings
$C^{(1,1,d-2)}_{a,b,c} : \dol{1}{1}(X)\times \dol{1}{1}(X)\times 
\dol{d-2}{d-2}(X) \rightarrow
{\mathbb C}$ are given by the period integrals 
of the holomorphic $d$-form on the
mirror $X^*$ as
\[ 
 C^{(1,1,d-2)}_{a,b,c}=\partial_{t_a} \partial_{t_b} \Pi^{(2)}_d
 \eta^{(2)}_{dc} \ .
\]
Here $t_a$ are the periods linear in logarithms\footnote{These 
are the flat coordinates on the complex structure moduli
space of $X^*$, i.e.\ the ones in which the Gauss-Manin connection on the
bundle $\cB \cong H^d(X^*,{\mathbb C})$ 
over this moduli space is flat. Under the
mirror map they become the flat coordinates on the moduli space of
the complexified K\"ahler structure of $X$.}
and
$\Pi^{(2)}_d$ are the quadratic ones, both in a gauge such
that the unique series solution for the periods around the large complex
structure point is equal to $1$.
On the other hand, $\eta^{(2)}_{dc}$ are purely topological 
two-point functions.
They define a metric on the primitive subspace of $\dol{2}{2}(X)$
by the cup product pairing on a fixed topological basis
$\{\cO^{(2)}_a\}$ of this subspace as
\beq
 \eta^{(2)}_{ab}=\langle \cO^{(2)}_a , \cO^{(2)}_b \rangle 
 =\int_X \cO^{(2)}_a \wedge \cO^{(2)}_b
 \ . \label{2pt:fun}
\eeq
Via mirror symmetry it coincides with the analogous cup product pairing
between B model observables on the mirror $X^*$, which are related
to the above basis elements $\cO^{(2)}_a$ by the mirror map.
Hence $\eta^{(2)}$ is a symmetric, invertible 
$\tilde{h}^{2,2}\times \tilde{h}^{2,2}$ matrix with integer entries. 
We denote its inverse by $\eta_{(2)}$. 

Our situation differs from that above in that we are interested in the point 
of the string moduli
space where an enhanced 
gauge symmetry arises, which is not the large radius point.
Furthermore we will take the field theory limit in which gravitational and
stringy modes decouple. 
In this limit the mirror $X^*$ turns into a complex surface $W_{\rm rig}$,
which we will call the local rigid surface. The attribute local refers to the
fact that it is determined by the local singularity structure of $X$, i.e.\
the fibration of exceptional divisors of the resolved $K3$ over the base
$\Bp^2$. In particular the number of twisted chiral superfields 
that do not 
decouple in this limit is given by the number $k$ 
of such exceptional divisors.
Moreover, as was established in \cite{selfd} for 3-folds, there
exists a map  $f: H_4(X^*,{\mathbb Z})\rightarrow 
H_2(W_{\rm rig},{\mathbb Z})$, such that
the subset of periods of the holomorphic 4-form $\Omega_{(4,0)}$ on $X^*$ that
form a closed monodromy problem by themselves is given 
in the rigid limit by periods of a
meromorphic 2-form $\la$ on $W_{\rm rig}$ as
$\left. \int_{\g} \Omega_{(4,0)}\right|_{\rm rigid}=\int_{f(\g)} \la$.
These periods give the scalar
components of the twisted chiral superfields not decoupling in the rigid limit
and of their magnetic duals.
In \cite{fi} it turned out that in the resulting $U(1)^k$ gauge theory
the r\^ole of the
flat coordinates $t_a$ is played by the $k$ series solutions 
$\s_a$ of the periods of $\la$ and that the $\Pi^{(2)}_d$ are replaced 
by the $k$ logarithmic solutions $\s_{D1,d}$.
The latter arise as periods over cycles that are the image under the map $f$
of 4-cycles in $X^*$, which are dual to 
observables of the B model on $X^*$ that correspond, via the mirror map, to
elements of the primitive subspace of $\dol{2}{2}(X)$.
Identifying 
the field theory limit of the exponential of
(\ref{Chern:Simons}) with
$\sim \exp(2 \pi i\tau_b(\s) n_b)$ we are hence led to
\beq
 \partial_{\s_a}\partial_{\s_b}\tW(\s)\sim \sum_{c}
  \nu_c \left. \langle \cO^{(1)}_{a}
  \cO^{(1)}_{b} \cO^{(2)}_{c} \rangle_{X}\right|_{\rm rigid}
  \sim \sum_{c,d} \nu_c 
  \partial_{\s_a}\partial_{\s_b}\s_{D1,d}(\s)\:
  \eta^{(2)}_{dc} \label{dW}
\eeq
or (modulo an additive constant and linear terms in $\s$)
\beq
 \tW(\s)\sim \sum_{c,d} \nu_c\: \eta^{(2)}_{cd} \,
  \s_{D1,d}(\s) \ . \label{W}
\eeq
The indices $a,b,c,d=1,\ldots,k$ correspond to the $k$ indices in 
$\{1,\ldots,h^{1,1}(X)\}$ and $\{1,\ldots,\tilde{h}^{2,2}(X)\}$, respectively,
which are related to periods surviving the field theory limit. 
Furthermore we kept the symbol $\eta^{(2)}$ for the resulting non-degenerate 
$k\times k$ submatrix of (\ref{2pt:fun}).

In view of (\ref{def:tau}) and (\ref{W}) the Fayet-Iliopoulos couplings
are given (modulo a possible additive constant) by
\beq
 \tau_a(\s) \sim \sum_{c,d=1}^{k} \nu_c\: \eta^{(2)}_{cd} \,
  \frac{\partial \s_{D1,d}}{\partial \s_{a}}
 \equiv \sum_{c,d=1}^{k} \nu_c\: \eta^{(2)}_{cd}\,
  \hat{\tau}_{ad}(\s) \ , \qquad a=1,\ldots,k\ .  \label{tau}
\eeq
The objects $\hat{\tau}_{ad}=\partial_{\s_{a}} \s_{D1,d}$ are very 
reminiscent of the gauge couplings 
$\tau_{ij}=\partial_{a_i} a_{D,j}$
of the $N=2$ supersymmetric
$SU(k+1)$ gauge theory in four dimensions. In that case the
rigid limit of the mirror Calabi-Yau 3-fold is a genus $k$ Riemann surface
$S$, whose $k$ holomorphic 1-forms $\omega_i$ stem from a meromorphic
1-form $\lambda_{\rm SW}$ as derivatives w.r.t.\ the $k$ moduli of the Coulomb
branch. Its $2k$ independent 1-cycles can be chosen to form a symplectic
basis $\{\alpha_j,\beta_j\}_{j=1}^{k}$ of $H_1(S,{\mathbb Z})$ such that
the gauge couplings are given in terms of the $k\times k$ parts
$A_{ij}=\int_{\alpha_j}\omega_i=\partial_{u_{i+1}}\int_{\alpha_j} 
 \lambda_{\rm SW}=\partial_{u_{i+1}} a_j$ and
$B_{ij}=\int_{\beta_j}\omega_i=\partial_{u_{i+1}}\int_{\beta_j} 
 \lambda_{\rm SW}=\partial_{u_{i+1}} a_{D,j}$
of the period matrix as $\tau_{ij}=(A^{-1}B)_{ij}$.

In complete analogy,
the local rigid surface $W_{\rm rig}$, which arises from the 
Calabi-Yau 4-fold we consider, is furnished with $k$ holomorphic 2-forms
$\omega_a$ stemming from a meromorphic 2-form $\lambda$ as derivatives 
w.r.t.\ the $k$ moduli. However, for even complex dimensional manifolds
the vanishing of $\int \Omega \wedge \Omega$ implies quadratic algebraic
dependences between the periods of the holomorphic form $\Omega$.
We therefore expect to find $3k$ independent 2-cycles
$\{\alpha_j,\beta_j,\gamma_j\}_{j=1}^{k}$ in $H_2(W_{\rm rig},{\mathbb Z})$,
such that 
the $\alpha$s intersect with $\gamma$s and the $\beta$s only among
themselves. Their $k\times 3k$ period matrix then comprises the $k\times k$
parts
\bea
A_{ab} &=& \int_{\alpha_b}\omega_a = \partial_{u_{a+1}}\int_{\alpha_b} 
 \lambda = \partial_{u_{a+1}} \s_b\ , \non \\
B_{ab} &=& \int_{\beta_b}\omega_a = \partial_{u_{a+1}}\int_{\beta_b} 
 \lambda = \partial_{u_{a+1}} \s_{D1,b}\ , \non \\
C_{ab} &=& \int_{\gamma_b}\omega_a = \partial_{u_{a+1}}\int_{\gamma_b} 
 \lambda = \partial_{u_{a+1}} \s_{D2,b} \ , 
 \label{period:matrix}
\eea
such that 
the couplings $\hat{\tau}_{ab}$, as defined in (\ref{tau}), are given by
$\hat{\tau}_{ab}=(A^{-1}B)_{ab}$. 
Whereas the $\s_b$ are series in the moduli, 
$\s_{D1,b}$ and $\s_{D2,b}$ are logarithmic and double
logarithmic, respectively, 
and this is precisely what
accounts for the logarithmic one-loop
correction in (\ref{tau:RG}).

In the following we will compute the periods $\s_a$, $\s_{D1,a}$ 
and $\s_{D2,a}$
using geometric engineering and
in a subsequent subsection (\ref{sec:yuk}) the four-point Yukawa couplings. 
This will allow us to exhibit the 
analogue of rigid special geometry for 4-folds.

\subsection{Geometric engineering}

The 
relevant part of the 
4-fold geometry is most efficiently described with the
help of toric geometry. Doing so, the Mori vectors describing the two
intersecting $\Bp^1$'s fibered over a base $\Bp^2$ (together with the
canonical line bundle that will be irrelevant for the purpose of this
section) are
\beq
\begin{array}{c@{\:=\: \left( \, \right.}*{6}{r@{\, , \,}}
r@{ \left. \,\right)}l} 
  l^{(1)} & 1 &-2 & 1 & 0 & 0 & 0 & 0 & \ , \\
  l^{(2)} & 0 & 1 &-2 & 1 & 0 & 0 & 0 & \ , \\
  l^{(3)} &-3 & 0 & 0 & 0 & 1 & 1 & 1 & \ .
\end{array} \label{mori:loc}
\eeq
The local mirror $X_{\rm loc}^*$ is then the complex surface given by
\beq
 X_{\rm loc}^* = \left\{ 0=P(y)=\sum_{i=1}^7 a_i y_i \right\} \ ,
\eeq
where the variables $\{y_i\}$ are projective and subject to the constraints
$1=\prod_{i=1}^7 y_i^{l_{i}^{(j)}}$, $j=1,2,3$. A solution to these is given by
\[
 (y_1,\ldots,y_7)=\left(s^3,s^2 t,s t^2,t^3,s^2 z,s^2 w,\frac{s^5}{z w}\right)
\]
with $[s,t,z,w]\in \Bp^3$, such that $P$ is a homogeneous 
Laurent polynomial of degree 3
\beq
 P=a_1 s^3+a_2 s^2 t+a_3 s t^2+a_4 t^3+a_5 s^2 z+a_6 s^2 w+a_7 
  \frac{s^5}{z w} \ . \label{mir:loc}
\eeq
In terms of the algebraic coordinates $a,b,c$
on the moduli space of complex structure
deformations of $X_{\rm loc}^*$, where
\beq
 a=\frac{a_1 a_3}{a_2^2} \ , \qquad b=\frac{a_2 a_4}{a_3^2} \ , \qquad
 c=\frac{a_5 a_6 a_7}{a_1^3} \ , \label{alg:var}
\eeq
the discriminant of $X_{\rm loc}^*$ reads
\beq
 \tilde{\D}=a b c \D=a b c ((\D_{\rm cl})^3+27 a^3 c\, q(a,b,c)) \ ,
 \label{dis:loc}
\eeq
with $q(a,b,c)$ a polynomial and
\[
 \D_{\rm cl}=-1+4 a +4 b-18 a b+27 a^2 b^2 \ .
\]
This discriminant is itself singular at the point $(a,b,c)=(1/3,1/3,0)$ and
expanding around this singularity as
\beq
 a = \frac{1}{3} - \left(\frac{1}{3}\right)^{2/3} \eps^2 u \ , \qquad  
 b = \frac{1}{3} - \left(\frac{1}{3}\right)^{2/3} \eps^2 u 
  +3 \eps^3 v  \ , \qquad
 c = \eps^9 \La^9 \ , \label{exp:locx}
\eeq
we find to lowest order in $\eps$
\beq
 \frac{1}{\eps^{18}}\D= -(4 u^3-27 v^2)^3-162 \La^9 v (4 u^3+9v^2)+27 \La^{18}
 +\cO(\eps) \ . \label{dis:field}
\eeq
Note that whereas for 3-folds the classical discriminant splits
quadratically, for 4-folds we have a cubic splitting \cite{fi}. 
In order to find
the correct variables on the field theory moduli space, we blow up
the singular point $(a,b,c)=(1/3,1/3,0)$ until we get divisors with only
normal crossings \cite{pointpart}. For one 
particular choice of coordinate patch this leads 
to the following variables
\bea
 z_1 &=& b-a = 3 \eps^3 v \ , \non \\
 z_2 &=& \frac{(a-\frac{1}{3})^3}{(b-a)^2} = -\frac{1}{81} 
 \frac{u^3}{v^2} \ , \non \\
 z_3 &=& \frac{c}{(b-a)^3} = \frac{1}{27} \frac{\La^9}{v^3} \ .
  \label{var:floc}
\eea
The sections $\s,\s_{D1},\s_{D2}$
obey a Picard-Fuchs
system of regular singular differential equations. In terms of the algebraic
coordinates $(a,b,c)$ this system takes the form
\bea
 \cL_1 &=& (\theta_a -3\theta_c)(\theta_a -2\theta_b)-
 a(-2\theta_a +\theta_b)(-2\theta_a+\theta_b-1) \ ,
 \non \\
 \cL_2 &=& \theta_b (-2\theta_a+\theta_b)-
 b(\theta_a-2\theta_b) (\theta_a-2\theta_b-1) \ , \non \\
 \cL_3 &=& \theta_c^3-c(\theta_a-3\theta_c)(\theta_a-3\theta_c-1)
 (\theta_a-3\theta_c-2) \ , 
 \label{pic:lstr} 
\eea
where $\theta_a=a \partial_a$, etc.
After transforming to the variables $z_1,z_2,z_3$, rescaling the periods 
$\pi_{\rm old}=\eps\pi_{\rm new}=\sqrt{\al^{\prime}}\pi_{\rm new}$, where 
$\eps\sim z_1^{1/3} z_3^{1/9}$, and taking the field theory limit 
$\eps \rightarrow 0$, we are left with two independent differential operators
\bea
 \cD_1 &=& \frac{1}{27} \theta_2 \left(\theta_2-\frac{1}{3}\right)+z_2
 \left(\frac{2}{3} \theta_2+\theta_3\right)\left(\frac{2}{3}
 \theta_2+\theta_3+\frac{1}{3}\right) \ , \non  \\
 \cD_2 &=&
 -\left(\theta_3+\frac{1}{9}\right)^3+
 z_3\left(\frac{2}{3}\theta_2+\theta_3\right)
 \left(\frac{2}{3}\theta_2+\theta_3+\frac{1}{3}\right)
 \left(\frac{2}{3}\theta_2+\theta_3+\frac{2}{3}\right) \ , \label{pic:field}
\eea
where $\theta_i=z_i \partial_{z_i}$.
Solutions to these are easily found, using the Frobenius method, by making the
ansatz
\beq
 \ts(s,t;z_2,z_3)=\sum_{n,m\geq 0} c(n,m;s,t)z_2^{n+s}z_3^{m+t} \ . 
 \label{ansatz}
\eeq
This determines the indices to be 
\beq
 (s,t) \in \left\{ \left(0,-\frac{1}{9}\right),\left( \frac{1}{3},-\frac{1}{9}
 \right) \right\} \label{ind}
\eeq
and, remembering that we chose the twisted chiral fields to have scaling
dimension $1$, we fix the first coefficient to be $c(0,0;s,t)=\La$. 
Recursion relations then imply that the general coefficient is given by
\bea
 c(n,m;s,t) &=& \La 
 \left\{ \prod_{i=1}^n \left[ (-27)\left(\frac{2}{3}(i-1+s)+t\right)
  \left(\frac{2}{3}(i-1+s)+t+\frac{1}{3}\right) \right] \right\} 
  \times \non \\
 & & \times
 \frac{(\frac{2}{3}(n+s)+t)_m (\frac{2}{3}(n+s)+t+\frac{1}{3})_m 
 (\frac{2}{3}(n+s)+t+\frac{2}{3})_m}{(s+1)_n (s+\frac{2}{3})_n
  [(t+\frac{10}{9})_m ]^3 } \ , \label{coeff}
\eea
where we have used the Pochhammer symbol 
$(a)_m = \G(a+m)/\G(a)=\prod_{i=0}^{m-1} (a+i)$.
Since the derivatives w.r.t.\ $s$ and $t$ 
commute with $\cD_1$ and $\cD_2$ and since
the first two derivatives w.r.t.\  $t$ of the indicial equations vanish at our
given pairs of indices, we find the following set of solutions in a 
neighbourhood
of $(z_2,z_3)\sim (0,0)$:
two series solutions
\beq
 \ts_i(z_2,z_3)=\left. \ts(s,t;z_2,z_3)\right|_{(s,t)_i} \ , \label{series}
\eeq
two logarithmic solutions
\bea
 \ts_{D1,i}(z_2,z_3) &=& \partial_t 
 \left. \ts(s,t;z_2,z_3)\right|_{(s,t)_i} \label{log} \\
  &=& \log(z_3) \ts_i(z_2,z_3)+\sum_{n,m\geq 0}\left(\partial_t c(n,m;s,t)
  \right) \left. z_2^{n+s} z_3^{m+t}\right|_{(s,t)_i} \non
\eea
and two double-logarithmic solutions
\beq
 \ts_{D2,i}(z_2,z_3) = \partial_t^2 
 \left. \ts(s,t;z_2,z_3)\right|_{(s,t)_i} 
 \ . \label{log2}
\eeq
The expansions for the series solutions read
\bea
 \ts_1(z_2,z_3) &=& \frac{\La}{z_3^{1/9}}\left(1+z_2-\frac{10}{729}z_3
 -4 z_2^2+\frac{440}{729}z_2 z_3 -\frac{1540}{531441}z_3^2 + 
 \frac{77}{3}z_2^3 - \right. \non \\
 & & \hspace{10mm} \left. - \, \frac{10472}{729}z_2^2 z_3 
 +\frac{261800}{531441}z_2 z_3^2-\frac{12042800}{10460353203}z_3^3
 +\ldots \right) \non \\
 \ts_2(z_2,z_3) &=& \frac{\La z_2^{1/3}}{z_3^{1/9}}
 \left(1-z_2+\frac{28}{729}z_3
 +5 z_2^2-\frac{910}{729}z_2 z_3 +\frac{7280}{531441}z_3^2 -
 \frac{104}{3}z_2^3 + \right. \non \\
 & & \hspace{10mm} \left. + \, \frac{19760}{729}z_2^2 z_3 
 -\frac{760760}{531441}z_2 z_3^2+\frac{76076000}{10460353203}z_3^3
 +\ldots \right) \label{exp:series} 
\eea
and for the logarithmic solutions as functions of the series solutions
\bea
 \ts_{D1,1}(\ts_1,\ts_2) &=& 9\; \ts_1 \log\left(\frac{\La}{\ts_1}\right)
 +\frac{27 \ts_2^9}{8 \ts_1^8} -\frac{27 \ts_2^6}{10 \ts_1^5}
 +\frac{9 \ts_2^3}{2 \ts_1^2}+\frac{55 \La^9 \ts_2^3}{27 \ts_1^{11}} - \non \\
 & & -\, \frac{11 \La^9}{243 \ts_1^8}-\frac{2443 \La^{18}}{354294 \ts_1^{17}} 
  +\ldots \non \\
  \ts_{D1,2}(\ts_1,\ts_2) &=& 9\; \ts_2 \log\left(\frac{\La}{\ts_1}\right)
 -\frac{27 \ts_2^{10}}{10 \ts_1^9} +\frac{27 \ts_2^7}{14 \ts_1^6}
 -\frac{9 \ts_2^4}{4 \ts_1^3}-\frac{5 \La^9 \ts_2^4}{\ts_1^{12}} + \non \\
 & & +\, \frac{59 \La^9 \ts_2}{243 \ts_1^9}
 +\frac{12157 \La^{18}\ts_2}{177147 \ts_1^{18}} 
 +\ldots  \label{exp:logs}
\eea
These sections have a geometric interpretation as period integrals of a
meromorphic 2-form $\la$ over 2-cycles in the rigid surface, which
arises from the local mirror in the 
field theory limit and which generalizes the
Seiberg-Witten curve arising from 3-folds.
The physical set of periods $\s_{b}$, $\s_{D1,b}$, $\s_{D2,b}$ is one that 
corresponds to 
an integral basis of 2-cycles with Weyl-invariant intersection form
such that the series solutions satisfy the
Casimir relations
\bea
\s_1^2(u,v)+\s_2^2(u,v)-\s_1(u,v) \s_2(u,v) &=& u+\ldots \ , \non \\
\s_1(u,v) \s_2(u,v) [\s_1(u,v)-\s_2(u,v)] &=& v+\ldots 
 \ , \label{Casimir}
\eea
where $\ldots$ indicate corrections that vanish in the classical limit 
(which in our coordinate patch (\ref{var:floc}) is at $v\rightarrow \infty$).
Using the notation 
$\s_{(0,b)}\equiv \s_b$, $\s_{(1,b)}\equiv \s_{D1,b}$ and 
$\s_{(2,b)}\equiv \s_{D2,b}$, 
such a set of periods is obtained from
the above solutions of the Picard-Fuchs equations by the 
linear transformation
\beq
 \s_{(a,b)}=c_a \sum_{c=1}^{2} M_{bc} \ts_{(a,c)} 
 \ , \qquad a=0,1,2  \label{mon:trafo}
\eeq
with
\beq
 M = \left(
 \begin{array}{cc}
  - (1-i\sqrt{3})/(2\; 3^{1/3}) & -1 \\
  - 1/(3^{1/3}) & (-1)^{2/3}
 \end{array} \right)  \label{trafo:M}
\eeq
and $c_0=1$, $c_1=i/(6 \pi)$ and $c_2=-1/(36 \pi^2)$.
The monodromy around $v\sim \infty$ for constant $u$ then acts on the
period vector $(\s_{1},\s_{2},\s_{D1,1},\s_{D1,2},\s_{D2,1},\s_{D2,2})^t$
as matrix multiplication by\footnote{
Actually the blocks below the diagonal could be altered by a change of
homology basis, which adds series solutions to logarithmic ones and
series and logarithmic solutions to double logarithmic ones 
and would
still meet our requirements. It is however important to notice that such
a change of basis leaves the relations (\ref{dW}) and (\ref{sg}) invariant
and reflects the indeterminacy mentioned above (\ref{W}).}
\beq
 M_v^{(\infty)}=\left( 
 \begin{array}{ccc}
   N & 0 & 0 \\
   N & N & 0 \\
   N & 2N & N
 \end{array} \right) \qquad   \mbox{where} \quad
 N=\left( 
 \begin{array}{cc}
   -1 & 1 \\
   -1 & 0
 \end{array} \right) \ . \label{mon:N}
\eeq
Note that $N^3$ is the identity matrix. 

For our specific example we can easily 
write down the equation for the local rigid
surface and the meromorphic 2-form $\la$. 
After shifting the $t$-variable such that the quadratic
term in $t$ disappears from (\ref{mir:loc}), and using (\ref{alg:var}) and
(\ref{exp:locx}), the local rigid surface takes the form
\[
 W_{\rm rig}=\left\{ z+w+\frac{\La^9}{z w}+P_{A_2}(x;u,v) =0 \right\} \ ,
\]
with $P_{A_2}(x;u,v)=x^3-u x-v$ the simple singularity of type $A_2$.
Equivalently we can take the polynomial form
\beq
 f_{\rm rig}=z^2 w+z w^2+\La^9+z w P_{A_2}(x;u,v) \ . \label{rig:surlo}
\eeq
As 
for the Seiberg-Witten curve,
this rigid surface is no longer a
Calabi-Yau space.
Rather, we expect
to find two holomorphic 2-forms 
that are derivatives of $\la$ w.r.t.\  $u$ and $v$, respectively. 
There are curves in the field theory moduli space 
(spanned by $u$ and $v$)
above which $\{f_{\rm rig}(w,x,z)=0\}$ is a singular space. Away from such
subvarieties, i.e.\ where 
$f_{\rm rig}$ defines a smooth surface, we can use the
Poincar\'e residue map to construct a holomorphic 2-form. 
Indeed, whenever we have 
$\left. \partial_w f_{\rm rig}\right|_{\{f_{\rm rig}=0\}} \equiv \sqrt{-4\La^9
z+z^2(z+P_{A_2})^2}\neq 0$ a 
suitable meromorphic 2-form is given by
\beq
 \la=-\frac{1}{z}\log\left[ 2z(z+P_{A_2})+2\sqrt{-4\La^9
  z+z^2(z+P_{A_2})^2}\, \right] dz\wedge dx \ , \label{lambda}
\eeq
such that
\[
 \om_1 = \partial_v \la = \frac{dz\wedge dx}{\sqrt{-4\La^9
  z+z^2(z+P_{A_2})^2}}= \frac{dz\wedge dx}{\partial_w f_{\rm rig}} 
\]
and
\[
 \om_2 = \partial_u \la = \frac{x\; dz\wedge dx}{\sqrt{-4\La^9
  z+z^2(z+P_{A_2})^2}} = \frac{x\; dz\wedge dx}{\partial_w f_{\rm rig}}
\]
are two holomorphic 2-forms, as we had expected.
In the case $\left. \partial_w f_{\rm rig}\right|_{\{f_{\rm rig}=0\}}=0$, but
$\left. \partial_z f_{\rm rig}\right|_{\{f_{\rm rig}=0\}} \equiv \sqrt{-4\La^9
w+w^2(w+P_{A_2})^2}\neq 0$, we just have to trade $z$ for $w$ in the 
above forms.

\subsection{The Yukawa couplings and rigid special geometry}
\label{sec:yuk}

In (\ref{dW}) we had found that the second derivatives of the twisted chiral 
potential are proportional to 
linear combinations of three-point Yukawa couplings of
the A model with target space the Calabi-Yau 4-fold $X$.
These couplings in turn were
identified with second derivatives of the logarithmic solutions 
to the period integrals of a meromorphic 2-form over 2-cycles in the
rigid surface.
In the previous subsection we used geometric engineering and
local mirror symmetry to compute these periods.
The purpose of this
subsection is to compute the four-point Yukawa couplings
directly from the Picard-Fuchs system.
This allows us to exhibit the structure
that generalizes rigid special geometry to 4-folds
in an example with two moduli.

The 4-fold we used above is a non-compact toric Calabi-Yau manifold, more
precisely the total space of a canonical line bundle. This non-compactness
did not bother us as long as we were only interested in 
properties of the rigid local mirror since, for these, only the compact
base space of the bundle was relevant. On the other hand, 
since the ring structure on the
cohomology of a non-compact manifold is not well defined, we 
need to use a
global model of the compact Calabi-Yau 4-fold $X$
in order to 
compute the Yukawa couplings (taking the 
field theory limit not until the end).
As already alluded to in the introduction, such a model is furnished,
for example, by 
the Fermat hypersurface of degree 36 in the resolution of the weighted
projective space $\Bp^5(18,12,3,1,1,1)$, which we called 
$X_{36}(18,12,3,1,1,1)$. This space is a fibration over $\Bp^2$ with fiber
a $K3$ surface given as $X_{12}(6,4,1,1)$ that is itself a fibration over
$\Bp^1$ with fiber this time a 2-torus $X_{6}(3,2,1)$. Note that this is the
same geometry as that of $X_{24}(12,8,2,1,1)$ 
\cite{k3fib,kava}, except that the base $\Bp^1$
of that $K3$-fibration has been traded for a $\Bp^2$ base. Since it is the
family of $K3$-fibers that determines the gauge symmetries, most of the
analysis is similar to the 3-fold case \cite{hoso1}.

The weighted projective space  $\Bp^5(18,12,3,1,1,1)$ is a toric variety 
whose fan comprises the one-dimensional cones
\[
\begin{array}{c@{\:=\:}l@{\ , \qquad}c@{\:=\:}l} 
 \nu_1 & (1,0,0,0,0) &
 \nu_2 & (0,1,0,0,0) \ , \\
 \nu_3 & (0,0,1,0,0) &
 \nu_4 & (0,0,0,1,0) \ , \\
 \nu_5 & (0,0,0,0,1) &
 \nu_6 & (-18,-12,-3,-1,-1) \ .
\end{array}
\]
They represent vertices of a reflexive polyhedron $\D$, which contains
in addition the origin $\nu_0$ as the only interior point, 
as well as the two integral vertices 
$\nu_7=(\nu_4+\nu_5+\nu_6)/3$ and
$\nu_8=(3\nu_3+\nu_4+\nu_5+\nu_6)/6$.
The latter represent exceptional divisors that are introduced in the
process of resolving the quotient singularities of the weighted projective
space. The enlarged vertices,
$\bar{\nu}_i=(1,\nu_i) \in {\mathbb Z}^{6}$,
satisfy three independent relations
$0=\sum_{i=0}^8 l_i^{(k)} \bar{\nu}_i$, $k=1,2,3$,
which define the Mori vectors
\beq
\begin{array}{c@{\:=\: \left( \, \right.}*{8}{r@{\, , \,}}
r@{ \left. \,\right)}l} 
  l^{(1)} &-6 & 3 & 2 & 0 & 0 & 0 & 0 & 0 & 1 & \ , \\
  l^{(2)} & 0 & 0 & 0 & 0 & 1 & 1 & 1 &-3 & 0 & \ , \\
  l^{(3)} & 0 & 0 & 0 & 1 & 0 & 0 & 0 & 1 &-2 & \ .
\end{array} \label{mori:glob}
\eeq
Applying standard techniques 
\cite{hoso1,baty,mondiv,hoso2} 
it follows that
the mirror $X^*$ is again a hypersurface in the same weighted projective
space $\Bp^5(18,12,3,1,1,1)$, given by
\bea
 p &=& a_0 z_1 z_2 z_3 z_4 z_5 z_6+a_1 z_1^2+a_2 z_2^3+a_3 z_3^{12}+
  a_4 z_4^{36}+a_5 z_5^{36}+a_6 z_6^{36} \non \\
 & & +\, a_7 (z_4 z_5 z_6)^{12}+a_8 (z_3 z_4 z_5 z_6)^{6} 
  \  . \label{mir:glob}
\eea
In terms of the algebraic coordinates
$a=a_1^3 a_2^2 a_8/a_0^6$, $b=a_4 a_5 a_6/a_7^3$,
$c=a_3 a_7/a_8^2$ 
on the moduli space of complex structure deformations of $X^*$ 
the Picard-Fuchs system for the periods $\pi_i(a,b,c)$ of the holomorphic
4-form on $X^*$ reads
\bea
 \cD_1 &=& \theta_a (\theta_a-2\theta_c)-12 a (6\theta_a+1)
 (6\theta_a+5) \ , \non \\
 \cD_2 &=& \theta_b^3-b (-3\theta_b+\theta_c)
 (-3\theta_b+\theta_c-1)(-3\theta_b+\theta_c-2) \ , \non \\
 \cD_3 &=& \theta_c (-3\theta_b+\theta_c)-
 c (\theta_a-2\theta_c)(\theta_a-2\theta_c-1) \ . \label{pic:globstr} 
\eea
From the analysis of the $X_{24}(12,8,2,1,1)$ model in \cite{k3fib} 
we can furthermore infer that the point of $SU(3)$ gauge symmetry 
enhancement is located in the string moduli space 
in rescaled variables $(x,y,z)=(432a,27b,4c)$
at $(x,y,z)=(\infty,0,1)$.
Expanding around this point as
\beq
 x = \frac{1}{2\eps^3 u^{3/2}} \ , \qquad 
 y = 3 \sqrt{3}\: \eps^9 \La^9   \ , \qquad
 z = 1-2\eps^3 u^{3/2} -3\sqrt{3}\: \eps^3 v \ , \label{exp:globx}
\eeq
we indeed find the discriminant of (\ref{pic:globstr}) 
to coincide with (\ref{dis:field}) at lowest order and up to 
an irrelevant factor.

Next, we compute the four-point Yukawa couplings following \cite{hoso2}.
This means that we first compute the four-point functions
of the B model on $X^*$ directly from the Picard-Fuchs system 
(\ref{pic:globstr}), 
transform them to the field theory variables and to the correct gauge
and take the rigid limit.
Whereas in \cite{hoso2} this gauge was chosen such that the unique fundamental
period around the large complex structure point was scaled to $1$, in our 
situation
it is fixed by the requirement that the Picard-Fuchs system 
(\ref{pic:globstr}), when transformed to this gauge, reduces 
in the field theory limit to the rigid
system (\ref{pic:field}). 
The rigid limit of the A model four-point Yukawa couplings is then 
obtained\footnote{These computations are straightforward but somehow
unhandy. Therefore we refer to the appendix for details and 
present the verification
of (\ref{sg}) only for one particular four-point coupling.}
by going to the flat coordinates $\s_i$.
As has already been mentioned, 
the four-point couplings are entirely determined by the two- and three-point 
functions as
\beq
  C_{\s_i \s_j \s_k \s_l} =
 \frac{1}{4!} \sum_{\pi \in S_4} \sum_{e,f=1}^{\tilde{h}^{2,2}} 
  C_{\s_{\pi(i)},\s_{\pi(j)},e}^{
  (1,1,2)} \; \eta_{(2)}^{ef} \; C_{\s_{\pi(k)},\s_{\pi(l)},f}^{
  (1,1,2)} \ . \label{yuk:fact} 
\eeq
Our identification of the A model couplings
$\left. C_{i,j,k}^{(1,1,2)}\right|_{\rm rigid} \sim 
\partial_{\s_i}\partial_{\s_j} \s_{D1,h} \eta^{(2)}_{hk}$ 
hence leads to the following relations
\bea 
  \left. C_{\s_i \s_j \s_k \s_l} \right|_{\rm rigid} 
 &=& \frac{\mbox{const}}{4!} \sum_{\pi \in S_4} \sum_{e,f,g,h} 
  \left(\partial_{\s_{\pi(i)}}\partial_{\s_{\pi(j)}} \s_{D1,g} \:
  \eta^{(2)}_{ge} \right)
  \eta_{(2)}^{ef}  \left(\partial_{\s_{\pi(k)}}\partial_{\s_{\pi(l)}}
  \s_{D1,h} \: \eta^{(2)}_{hf} \right)  \non \\
 &=&  \frac{\mbox{const}}{4!} \sum_{\pi \in S_4} \sum_{e,f} 
  \left(\partial_{\s_{\pi(i)}}\partial_{\s_{\pi(j)}} \s_{D1,e} 
   \right)
  \eta^{(2)}_{ef}  \left(\partial_{\s_{\pi(k)}}\partial_{\s_{\pi(l)}}
  \s_{D1,f} \right) \  . \label{sg}
\eea
Here $\eta^{(2)}$ is the non-degenerate
submatrix of the intersection form (\ref{2pt:fun}) 
on the primitive subspace of $\dol{2}{2}(X)$
corresponding to periods
that survive the rigid limit.
It is a symmetric, invertible $2\times 2$ matrix with integer
coefficients, whose inverse we denote by $\eta_{(2)}$.
The constant of proportionality 
appearing in (\ref{sg}) is of course the same for 
all couplings.
Having calculated the four-point functions and the periods, we can indeed match
the left and right sides of (\ref{sg}), which fixes the constant of 
proportionality and moreover 
determines the intersection form $\eta^{(2)}$ to be
the Cartan matrix of $SU(3)$
\beq
 \eta^{(2)}= \left(
 \begin{array}{cc} 2 & -1 \\ -1 & 2 \end{array} \right) \ . \label{eta}
\eeq
Relations (\ref{W}) and (\ref{sg}) represent the analogue of
rigid special geometry for 4-folds with several moduli, in the same
sense that special geometry for 3-folds manifests itself in the
relations $F=\frac{1}{2}Z^i F_i(Z)$ for the prepotential $F$, where
$Z^i$ and $F_i$ are periods of the holomorphic 3-form w.r.t.\  a
symplectic basis of $H_3(X_3,{\mathbb Z})$ and 
$C_{ijk}=\partial_i \partial_j \partial_k F$ for the three-point 
Yukawa couplings,
where the derivatives are w.r.t.\  the special projective coordinates 
$Z^i$. They are the rigid limit of the structure found in \cite{mayr4}
for the non-rigid case.

\section{Conclusion}
We have investigated the field theory limit of type IIA string
compactification on a 
Calabi-Yau 4-fold whose relevant part for the purpose of extracting field
theoretic properties consists of two intersecting $\Bp^1$'s fibered over a
common base $\Bp^2$. The rigid limit of the local mirror is a complex
surface that generalizes the Seiberg-Witten curve and on which there
exist two holomorphic 2-forms that stem from the same meromorphic
2-form as derivatives w.r.t.\ the two moduli.

The effective field theory that is the appropriate description in the
infrared is an $N=(2,2)$ supersymmetric gauge theory 
in two dimensions with abelian gauge
group $U(1)^2$. Its twisted chiral potential is of the form
$\tW=\nu\cdot\eta^{(2)}\cdot\s_{D1}$, where $\nu$ is a vector of 4-fluxes,
$\eta^{(2)}$ an intersection form and $\s_{D1}$ a period vector.
By explicit computation of the period integrals 
as solutions of the Picard-Fuchs equations
and of the four-point Yukawa
couplings we were able to exhibit the generalization of rigid special 
geometry to 4-folds in a non-trivial example with two moduli.
This structure manifests itself in the relation 
$C_{(4)}=(\partial^2 \s_{D1}) \cdot \eta^{(2)} \cdot (\partial^2 \s_{D1})$
between these four-point functions and derivatives of the 
middle periods w.r.t.\ 
flat coordinates.

We briefly mention a number of conclusions about two-dimensional gauge
theories as derived from type IIA compactifications on 4-folds, which
have already been discussed in \cite{fi} but apply to our example as well.
The major novelty of 4-folds is the r\^ole 
played by the primitive subspace
of $\dol{2}{2}(X)$ and its dual 4-cycles, respectively. They lead to new
discrete moduli of the gauge theories in two dimensions, the 4-fluxes.
If they all vanish, the theory just exhibits a
non-trivial K\"ahler potential. But once the 4-fluxes are switched on,
the structure of the theory becomes richer, as a twisted chiral potential,
FI couplings and a scalar potential are generated. Generically the last
seems to break supersymmetry, as was the case in the one-modulus 
example of   
\cite{fi}. The coefficient of the logarithmic term in the 
FI couplings can furthermore be
interpreted as indicating the presence of massive chiral matter we
had not accounted for in the geometrical set-up.
Finally, the choice of base is not unambiguous, in contrast to the case
of 3-folds, but the instanton series depends on this choice. We do not
know how to interpret or resolve this ambiguity.

\begin{center}
{\bf Acknowledgements}
\end{center}
I would like to thank Ulrike Feichtinger, Peter Mayr and
especially Wolfgang Lerche
for discussions and sharing their insight.

\appendix
\section{The four-point Yukawa couplings}
In this appendix we discuss some details of the steps occuring in our
computation of the rigid limit of the four-point Yukawa couplings, the
general idea of which has already been explained above (\ref{yuk:fact}).
Subsequently we present the verification of (\ref{sg}) 
for one particular coupling.

First we compute\footnote{These
computations were performed with the help of the program Lop4f.m, written
by Albrecht Klemm.}
the four-point Yukawa couplings
of the B model on
$X^*$, which is the mirror to
$X_{36}(18,12,3,1,1,1)$.
We write them in the rescaled algebraic variables $(x,y,z)$ and in order to
shorten the exposition we introduce the following abbreviations for the
components of the discriminant of (\ref{pic:globstr})
\[
 \D = (1+y) ((-1+z)^3+y z^3) ((-(-1+x)^2+x^2 z)^3+x^6 y z^3) \equiv 
 \D_1 \D_2 \D_3 \ .
\]
Furthermore $C^{(4,0,0)}$ stands for $C_{xxxx}$, $C^{(2,1,1)}$ for $C_{xxyz}$
and so forth.
\bean
 C^{(4,0,0)} &=& \frac{(-1+x)^2}{x^4 \D_3} \ , \qquad
 C^{(3,0,1)} = -\frac{(-1+x)^3}{2 x^3 z \D_3} \ , \\
 C^{(2,0,2)} &=& \frac{(-1+x)^4}{4x^2 z^2 \D_3} \ , \qquad
 C^{(1,0,3)} = -\frac{(-1+x)^5}{8 x z^3 \D_3} \ , \\
 C^{(0,0,4)} &=& -\frac{(-1+2x)}{16 z^3 \D_2 \D_3} \left[(3-3z+(1+y)z^2-6x^3
  \left( 2-3z+(1+y)z^2 \right)- \right.\\
 & & \hspace{15mm} -\, 4x\left( 3-3z+(1+y)z^2)+3x^4(1-2z+(1+y)z^2 \right)+ \\
 & & \hspace{15mm} +\, \left. x^2 \left(18-21z+7(1+y)z^2 \right) \right] \ , \\
 C^{(3,1,0)} &=& \frac{(-1+x)\left( -1+2x+x^2(-1+z)\right)}{6x^3 y \D_3} \ , \\
 C^{(2,1,1)} &=& -\, \frac{(-1+x)^2 \left( -1+2x+x^2(-1+z)\right)}{
  12x^2 y z \D_3} \ , \\
 C^{(1,1,2)} &=& \frac{(-1+x)^3 \left( -1+2x+x^2(-1+z) \right)}{
  24x y z^2 \D_3} \ , \\
 C^{(0,1,3)} &=& \frac{(-1+2x)}{48 y z^2 \D_2 \D_3} \left[ -2+3z-(1+y)z^2
  +4x \left(2-3z+(1+y)z^2 \right) - \right.  \\
 & & \hspace{15mm} -\, 2x^3 \left( -4+9z-3(2+y)z^2+(1+y)z^3 \right) + \\
 & & \hspace{15mm} +\, x^2 \left( -12+21z-(10+7y)z^2+(1+y)z^3 \right) + \\
 & & \hspace{15mm} +\, x^4 \left. \left( -2+6z-3(2+y)z^2+2(1+y)z^3 \right) 
  \right] \ , \\
 C^{(2,2,0)} &=& \frac{\left( -1+2x+x^2(-1+z) \right)^2}{36x^2 y^2 \D_3} \ , \\
 C^{(1,2,1)} &=& -\, \frac{(-1+x) \left( -1+2x+x^2(-1+z) \right)^2}{
  72x y^2 z \D_3} \ , \\
 C^{(0,2,2)} &=& -\, \frac{(-1+2x)}{144 y^2 z \D_2 \D_3} \left[ 1-2z+(1+y)z^2
  -4x \left( 1-2z+(1+y)z^2 \right) + \right.\\
 & & \hspace{15mm} +\, x^2 \left( 6-14z+(10+7y)z^2-2(1+y)z^3 \right) + \\
 & & \hspace{15mm} +\, x^4(-1+z) \left( -1+3z-3(1+y)z^2+(1+y)z^3 \right) + \\
 & & \hspace{15mm} +\, \left. 2x^3 \left( -2+6z-3(2+y)z^2+2(1+y)z^3 \right) 
  \right] \ , \\
 C^{(1,3,0)} &=& -\, \frac{(-1+x) \left( 1-4x-3x^2(-2+z)+x^3(-4+6z)+
  x^4(1-3z+3z^2) \right)}{216 x y^2 \D_1 \D_3} \ , \\
 C^{(0,3,1)} &=& -\, \frac{(-1+2x) \left( 1-4x+x^2(7-3z)+6x^3(-1+z)+
  3x^4(-1+z)^2 \right) z^2}{432 y^2 \D_2 \D_3} \ , \\
 C^{(0,4,0)} &=& -\, \frac{(-1+2x)z}{1296 y^3 \D_1 \D_2 \D_3} 
  \left[ -1+3z+(-2+y)z^2 - \right. \\
 & & \hspace{15mm} -\, 4x \left( -1+3z+(-2+y)z^2 \right) + \\ 
 & & \hspace{15mm} +\, x^2 \left( -6+21z+(-20+7y)z^2+(5-4y)z^3 \right) + \\
 & & \hspace{15mm} +\, 2x^3 \left( 2-9z-3(-4+y)z^2+(-5+4y)z^3 \right) + \\ 
 & & \hspace{15mm} +\, x^4 \left. \left( -1+6z+3(-4+y)z^2 
  - 2(-5+4y)z^3+(-3+6y)z^4 \right) \right] \ .
\eean
Using the transformation property 
\[
 C_{w_i w_j w_k w_l}=\sum_{m,n,p,q} \frac{\partial z_m}{\partial w_i}\, 
  \frac{\partial z_n}{\partial w_j}\, 
  \frac{\partial z_p}{\partial w_k}\, 
  \frac{\partial z_q}{\partial w_l}\, C_{z_m z_n z_p z_q}
\]
of the four-point Yukawa couplings and (\ref{exp:globx}) we transform
the above couplings to the field theory variables $(u,v)$.
We make sure that we use corresponding gauges for the holomorphic 4-forms 
in the local and global construction by rescaling the global periods by a
holomorphic function $1/f(u,v)$, such that the Picard-Fuchs system 
(\ref{pic:globstr}), when transformed to this gauge, reduces 
in the field theory limit to the rigid
system (\ref{pic:field}). This implies that we have to rescale the
Yukawa couplings by $1/f^2(u,v)$ after which we can take the rigid limit.

Next we have to change from variables $(u,v)$ to $(\s_1,\s_2)$. Since we
are going to compare leading terms in series expansions, it is easiest
to work with $\ts$ and $\ts_{D1}$, which are given in (\ref{exp:series}) and
(\ref{exp:logs}) and are related to the true $\s$ and $\s_{D1}$ by the
transformations (\ref{mon:trafo}), and to express everything in the
variables $(z_2,z_3)$ as given in (\ref{var:floc}).
For (\ref{sg}) this implies that we replace $(\s,\s_{D1})$ by 
$(\ts,\ts_{D1})$ whereby $\eta^{(2)}$ gets transformed to
$\tilde{\eta}^{(2)}=c_1^2 M^t \eta^{(2)} M$ with $c_1$ and $M$ as in
(\ref{trafo:M}).

Let us verify (\ref{sg}) for 
$C_{\ts_1 \ts_1 \ts_1 \ts_1}$. Performing
the above mentioned steps we find
\bean
 (-1)^{7/6}\frac{8}{3} C_{\ts_1 \ts_1 \ts_1 \ts_1} &=&
 z_2^{1/3}z_3^{2/9} \left( -162+1620 z_2-\frac{4184}{9} z_3 
 - 18306 z_2^2 +\frac{242000}{9} z_2 z_3 - \right.  \\
 & & \hspace{20mm}  
 -\, \frac{4584286}{6561} z_3^2 
 + 212652 z_2^3 -\frac{7880872}{9} z_2^2 z_3 +  \\
 & & \hspace{20mm} \left. +\, \frac{670041316}{6561} z_2 z_3^2 
 -\frac{117589832812}{129140163} z_3^3
 +\ldots \right) \ .
\eean
From (\ref{exp:series}) and (\ref{exp:logs}) we have
\bean
 \partial_{\ts_1}^{2} \ts_{D1,1} &=&
 z_3^{1/9} \left( -9+36 z_2-\frac{274}{81} z_3 -315 z_2^2 
 +\frac{25252}{81} z_2 z_3 -\frac{152633}{59049} z_3^2 \right.  \\
 & & \hspace{5mm} \left. +\, 3120 z_2^3 -\frac{890726}{81} z_2^2 z_3 
 +\frac{42870113}{59049} z_2 z_3^2 -\frac{2555818340}{1162261467} z_3^3
 +\ldots \right)
\eean
and
\bean
 \partial_{\ts_1}^{2} \ts_{D1,2} &=&
 z_2^{1/3} z_3^{1/9} \left( 9-54 z_2-\frac{202}{9} z_3 + 486 z_2^2
 -1072 z_2 z_3 +\frac{547126}{19683} z_3^2 -4860 z_2^3 + \right.  \\
 & & \hspace{20mm} \left. +\, 30524 z_2^2 z_3 -\frac{23885459}{6561} z_2 z_3^2
 +\frac{1355705192}{43046721} z_3^3 +\ldots \right) \ .
\eean
Thus (\ref{sg}) holds for $C_{\ts_1 \ts_1 \ts_1 \ts_1}$ if
\[
 \tilde{\eta}^{(2)}=\left(
 \begin{array}{cc} 0 & 1 \\ 1 & 0 \end{array} \right) \ .
\]
This implies the validity of (\ref{sg}) for $C_{\s_1 \s_1 \s_1 \s_1}$ with
$\eta^{(2)}$ the Cartan matrix of $SU(3)$ as stated in (\ref{eta}).
Using the same method one easily checks
that the same constant of proportionality  and the 
same intersection form $\eta^{(2)}$ work for all four-point Yukawa 
couplings, thus verifying (\ref{sg}).


\end{document}